# Laser Stimulation of Low-Energy Nuclear Reactions in Deuterated Palladium

K.P.Sinha[a] and A.Meulenberg[b†]

[a] Department of Physics, IISc, Bangalore 560012, India (kpsinha@gmail.com)
[b] Visiting Scientist - Department of Instrumentation, IISc, Bangalore 560012, India

Abstract

Models to account for the observed experimental results for low-energy nuclear reactions in palladium-deuteride systems are presented along with calculated results. The crucial idea is a mechanism of improved probability for the needed penetration of the Coulomb barrier for a D-D reaction. This facilitation occurs, in general, with the formation of $D^-$ ions at special frequency modes (e.g. via phonons) and, specifically for the laser-stimulated case, with utilization of enhanced optical potential at a selected interface. Both mechanisms may work individually, or together, to increase the probability of barrier penetration.

Introduction

Observations of anomalies in metal deuterides, such as excess heat and nuclear emissions, have led to the emergence of the field of low-energy nuclear reaction (LENR) in a solid matrix. Several groups have carried out careful investigations of this phenomenon and the general literature on the subject has grown immensely over the last 15 years, with thousands of papers covering various aspects of the results[1]. The LENR takes place in a heavily-deuterated solid matrix (such as Pd, Ti, Ni or some ceramic oxides)[2]. The host matrix provides interstitials, voids, crevices or channels for light atoms (such as D or H). Further-increased loading, by D(H), leads to the displacement of host metal atoms resulting in vacant host sites. For example, in the palladium-deuteride [Pd (D/H)] system, it is found that deformed lattices produce channels through which $D^+(H^+)$ have considerable mobility. Furthermore, there are defect regions in the matrix where these ions are subject to lattice forces and interactions with high-frequency modes (phonons or plasmons).

Owing to these phonon interactions, the existence of $D^-(H^-)$ ions at some locations in the matrix becomes favorable[3]. In fact, it has been shown theoretically that, even without the phonon interaction, the two-electron bound state occurs on hydrogen atoms in metals. This permits $H^-$ (or $D^-$) ions to exist (with the electron pair in a 1s bound state).[4] On addition of the electron-phonon interaction, these structures are further stabilized (relative to the single electron case). This interaction also increases the effective electron mass of singly-bound electrons; but, the effect is greater for the paired electrons.

As a result of refined experiments, many laboratories of the world have reported reliable and repeatable LENR in $PdD_x$ systems. Some workers have discovered hot spots in the systems suggesting the appearance of excess heat at some special points (regions) of the system (reported in IEEE spectrum, September 2004 pages 22-26). In this context, it is claimed by Letts and Cravens[5] and by Violante et. al.[6], that the laser stimulation of LENR in deuterated palladium is reproducible. A theoretical understanding of this effect has not yet been provided.

---

[†] Corresponding author (mules33@excite.com)

In what follows, we present a brief account of the experimental situation leading to these two effects (hot spots and laser stimulation) and then propose a mechanism, which, we believe, is responsible for the observations.

Experimental background and proposed theoretical concepts

As described by Letts and Cravens[5], a palladium cathode is fabricated by sequentially cold-working, polishing, etching and annealing prior to being electrolyzed in heavy water. The electrolytic loading with deuterium is carried out by placing the cathode in a magnetic field of 350 Gauss. After this is accomplished, Gold is co-deposited electrolytically on the cathode to yield a visible coating. Then the cathode is stimulated with a low-power laser (having a maximum power of 30 milliwatts at 661.5nm). It is found that the thermal response of the cathode reaches around 500 mW. (Maximum reported output was near 1.0 Watt.) The effect is repeatable (in their and in another laboratory), provided their method of sample preparation and experimental procedure is followed.

The other experiment[6] with laser stimulation of LENR in deuterated palladium is built up in a somewhat different manner. Violante et. al. confirm the light-polarization effect reported by Letts and Cravens and propose a mechanism to explain it. As a consequence, (to increase the optical coupling to the longitudinal phonons) an acid etch of the palladium cathode is used, to produce the roughened surface, but no thin gold plating is added. While the level of laser-induced power enhancement appears to be lower in Violante's work, the enhanced power has been measured over more than a hundred hours. The results thus indicate another aspect of the laser mechanism which could become important in continuous-operation LENR systems.

A central concept involved in the laser stimulation of LENR and generation of heat is the production of "hot sites" (perhaps in the surface contact between PdD nanoparticles and gold particles). It is suggested herein (as in Reference 6) that the activity of these regions is due to the excitation of surface modes (plasmons or polaritons[7]). The interparticle optical potential in the gap between $PdD_x$ and gold particles can change sign as the distance between particles decreases to the near-field region[8]. This also induces a force that pulls the particles together. In the next section, theoretical formulations and some numerical results are presented.

Theoretical Formulation and Results

The important points to note in the formulation of laser-stimulated LENR in the system described above are the following:

(1) The defects in the lattice structure of Pd on loading with D lead to the formation of crevices, voids, and channels in the matrix. While this process could be harmful in that it may prevent the lattice from becoming adequately loaded, it could be an essential ingredient in the creation of active sites in the Pd matrix.

(2) The interaction of electrons and nuclei with the optical-phonon modes can lead to the localization of electron pairs (bosons) on some deuterons with the formation of bosonic ions $D^-$ and the subsequent screening of nuclear charge.[3]

(3) The collective motion of the deuterons (driven by the phonons) can introduce "breathing" modes in the Pd lattice. These breathing modes, if they exist and if they are resonant with the deuteron motion, have three benefits. A) First, by enhancing deuteron migration (diffusion), they help to prevent premature build up of lattice-disruptive, non-uniform, deuteron densities. B) Second, enhanced diffusion of the deuterons allows rapid refilling and regeneration of the active sites, if they are not damaged by the D-D reaction. C) Third, (in the case of non-defect-site related D-D reaction) if resonant with the anti-phase mode of the deuterons, the Pd atoms could move apart as the deuterons come together. This separation not only permits direct collision of the adjacent interstitial deuterons, it provides an electron cloud to help screen the deuteron's repulsive Coulomb fields.

(4) The enhancement of the local electric field of the laser $E_L$, as compared to the incident electric field and due to the response of the material surface, constitutes the electrodynamical environment surrounding the ion pairs in question ($D^-$-$D^+$).

Bosonic Ions

The distortion of the local lattice around some sites occupied by D, along with its interaction with high-frequency modes (optical phonons, surface plasmons etc), reduces the electronic repulsion for two electrons on the same deuterium ion ($D^-$).[3,9] As a result, the normal Coulomb repulsion (U) between two electrons on the same orbital state of a $D^-$ ion is reduced to

$$U^* = U - 2E_d \qquad \text{where,} \qquad (1)$$

$$E_d = (g^*)^2 \hbar \omega_d, \qquad (2)$$

$$(g^*)^2 = g^2 \coth(\hbar \omega_d / 2 k_B T), \qquad (3)$$

g is the dimensionless electron-vibration coupling at temperature T=0, $\omega_d$ is the frequency of the mode, $\hbar$ and $k_B$ are the usual constants. When $U < 2E_d$, the electron pairs (in the singlet state) form composite bosons. In these sites, the state of $D^-$ with double occupancy will be more stable than would be a singly-occupied atomic state in the same site. Thus, the deuterium atoms will be distributed in the systems as $D^-$ in some sites (having local distortion in interaction with high frequency modes). The distortion of the lattice and the interaction with high-frequency modes make $U^*$ negative, hence leading to stability of the bosonic ion $D^-$ at the sites in question.[3,9]

The consequences of the bosonic ion are quite dramatic. First of all, the normally-considered monopole-monopole repulsion of the $D^+$-$D^+$ interaction (when both deuterons are ionized) is reversed (Figure 1) in the far-field. This means that, until the electron screen is penetrated significantly, the elements of the $D^+$-$D^-$ pair are attracted to each other (monopole-monopole). The screening only begins at the edge of the electron pair distribution. In free space, this monopole-monopole attraction makes any dipole-monopole and dipole-dipole effects, insignificant.

However, as a second effect (shown in Reference 3), the $D^+$-$D^-$ equilibrium position in a metal lattice is much closer than that of free molecular deuterium $D_2$ as a consequence of the increased effective electron mass from phonon interaction. This increased mass reduces the electron-distribution size (into the sub-nanometer range) and therefore the point at which the monopole-monopole attraction (of the $D^+$-$D^-$ pair) begins to diminish.

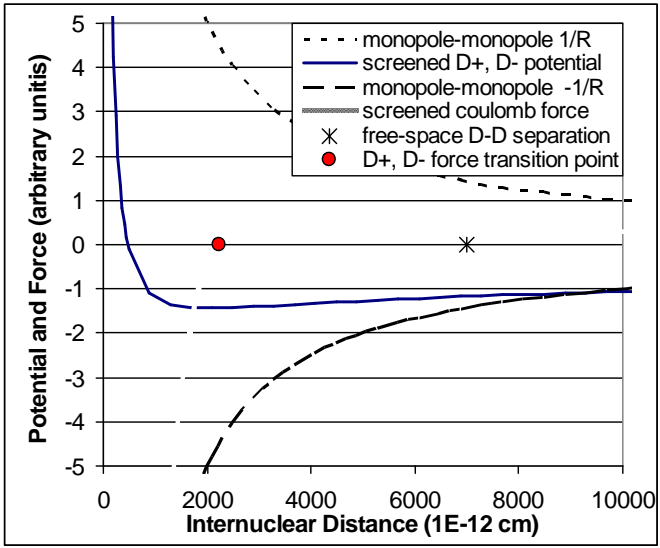

*Figure 1 Representative potentials and force associated with $D^+$-$D^-$ pairs. The transition point from attractive to repulsive force (zero-force point) is identified along with the free-space molecular-deuterium bond length.*

Qualitatively, as the $D^+$ deeply penetrates the phonon-modified electron screen of the $D^-$, it will see an abrupt increase in the positive charge.[3] [Another scenario, assuming that an electron is confined to oscillate between the deuterons, counters this expectation and proposes a deep (many keV) secondary attractive-potential well.][10] Even if the electrons are not highly localized, but (as a bosonic pair) switch from one deuteron to the other deuteron, there will still be a greatly-reduced Coulombic repulsion between the deuterons in this range. In this context, we can envisage a new kind of chemical bond (within a solid matrix) which involves the exchange of a real-space electron pair between the two $D^+$. This real-space electron pairing can bring the two deuterons into closer proximity than would otherwise be possible.

Ignoring 2nd order effects, we can see important consequences from the zero-force positions in Figure 1. The paired $D^-$ - $D^+$ system has a much reduced zero-force distance (~ 2 nm) relative to that of a $D_2$ molecule (~ 7 nm). It also has a different force dependence on displacement from this position. The restoring force, for $D^-$ - $D^+$ displacements at room temperature is not symmetric. It is weak and nearly uniform on the outward path relative to that on the inside of the zero-force position. This means that the recoil path is long relative to the average penetration path (e.g. 10 nm vs 1 nm). Thus, even if the closely-proximate deuterons were to disrupt the phonon-electron interaction, the deuterons would have achieved a higher impact velocity and a lower repulsive-field impact diameter than that calculated without the phonon interaction.

How significant are these two effects (Coulomb attraction and phonon-modified electron-screen dimensions)? For paired $D^-$ - $D^+$ systems in a crevice, void, channel or surface assuming an average pair separation (e.g., 7 nm), the mutual force of attraction is:

$$F = e_1(D^-)\, e_2(D^+) / r^2 \approx 23 \times 10^{-20} / 49 \times 10^{-18} \approx 5 \times 10^{-3} \text{ dynes}, \qquad (4)$$

which, using center-of-mass coordinates, can be equated to $F = (m_d/2)\, a$, where $m_r = m_d/2$ is the reduced deuteron mass and 'a' the relative acceleration between the deuterons. With $m_d \sim 3.3 \times 10^{-24}$ gms, $a = 3 \times 10^{21}$ cm sec$^{-2}$. Assuming that screening only prevents the monopole-monopole attraction from greatly increasing as the deuterons approach into the deep-screening region (~ 1 nm), we can estimate an acceleration path of ~10 nm. The relative velocity increases from an initial velocity of $5 \times 10^5$ cm sec$^{-1}$ (with this new acceleration and path length) to an impact velocity of $v_{d(impact)} = \sim 6.5 \times 10^6$ cm sec$^{-1}$. This gives a corresponding kinetic energy of $E_d = \frac{1}{2}(m_d v_d^2) = \sim 7 \times 10^{-11}$ ergs.

To determine the effect of this increased impact velocity, we estimate the penetration coefficient for the Coulomb barrier. The probability of penetration per D$^-$- D$^+$ interaction is given by [11]

$$P_T = \exp(-G), \quad (5)$$

for the very-low-energy regime and with the Gamow factor

$$G = e^2 / \hbar v_d. \text{ (Note that } v_d \text{ is parallel velocity, not frequency } \nu.) \quad (6)$$

First we study the dark case and then (in the next section) we examine the effect of laser stimulation. The cross section of a D$^-$ - D$^+$ reaction will involve the following situation. The nuclear forces of attraction operate only at short distances ($10^{-13} < R < 10^{-12}$ cm). Outside this range, the nuclei will feel the repulsive Coulomb potential as the two deuterons approach each other (up to $2R = \sim 10^{-12}$ cm). Owing to the electron cloud around D$^-$, the D$^+$ will penetrate the negative charge cloud (to which it is attracted) until it experiences the repulsive nuclear-Coulomb barrier deep into the screening region. The interaction cross-section $\sigma(D^-, D^+)$ is given by [9]

$$\sigma(D^-, D^+) = (\pi/k^2) P_T = (\pi/k^2) \exp(-e^2/\hbar v_d) \text{ cm}^2, \text{ where} \quad (7)$$

$$k^2 = (2 m_r E_d / \hbar^2) = m_r^2 v_d^2 / \hbar^2 \quad (8)$$

Using $m_r \sim 1.67 \times 10^{-24}$ gms, $v_{d\ (impact)} = \sim 6.5 \times 10^6$ cm sec$^{-1}$, $e = 4.8 \times 10^{-10}$ cgs units, and $\hbar = 10^{-27}$ erg sec, we have $\sigma(D^-, D^+) = (\pi/3.68 \times 10^{20}) \exp(-35.5) \text{ cm}^2 = \sim 3 \times 10^{-36} \text{ cm}^2$.

From a D-D calculation, at $v_{d\ (impact)} = 5 \times 10^5$ cm sec$^{-1}$, the value for $\sigma(D^-, D^+)$ is vanishingly small. Thus, the increase in impact velocity, in the range of interest (from $5 \times 10^5$ cm/sec to $6.5 +/- 2 \times 10^6$ cm/sec in Figure 2), makes a dramatic difference. Nevertheless, the coulomb barrier (because of its scattering potential) is still sufficient to prevent nuclear interaction of the deuterons. The neutral-particle cross-section for the nuclear-coalescence reaction (also shown in Figure 2) is $\pi (2R)^2 = \sim 3 \times 10^{-24}$ cm$^2$.

Unfortunately, beyond the enhanced-impact-velocity range calculated, the interaction cross-section does not increase as rapidly. It would take another order-of-magnitude increase in impact velocity for the deuterons to have a reasonable probability of penetrating to the critical nuclear-interaction radius (theoretical nuclear-coalescence interaction area).

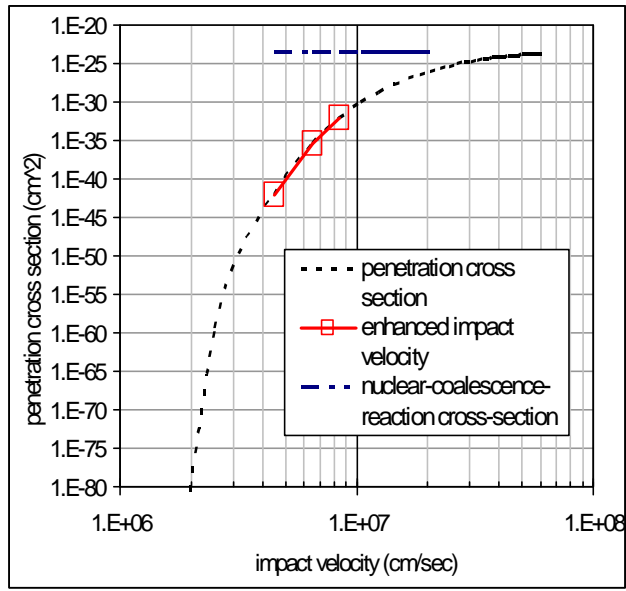

*Figure 2 Penetration cross section as a function of deuteron impact velocity. Enhanced velocity regime and nuclear-coalescence interaction area are shown for comparison.*

Thus, despite the higher impact velocity calculated with the $D^-$-$D^+$ attraction, the reaction is not likely under conditions of the calculation. However, additional effects may be considered. Electrons in the 1s orbital have a non-zero screening amplitude even at the nucleus, the effective charge of the proximate nuclei, may be less than 'e' i.e., say about 0.5e, or even 0.35e.[3] Then, the Gamow factor (equation 6) decreases by a factor of 2 – 3. Values, with the new factor G', for the expression $P_T$ = Exp [-G'] become Exp [-35.5/2] = ~$2 \times 10^{-8}$, or Exp [-35.5/3] = ~ $7 \times 10^{-6}$. This will give $\sigma(D^-,D^+)$ a value of the order $6 \times 10^{-28}$ to $10^{-25}$ cm$^2$. These larger values <u>are</u> approaching the high-probability range for the D-D interaction.

The paired $D^-$ - $D^+$ effect, described above, will increase the average impact velocity by about an order of magnitude. This brings the interaction cross section to within ~12±5 orders of magnitude of the nuclear-coalescence interaction area. Inclusion of effective-charge reduction from electron screening raises the cross section by another 7-10 orders of magnitude. If all occupied interstitial deuteron sites were possible nuclear-interaction sites, then fusion would be observed, under these conditions, even at low-loading levels (e.g., 1%). Thus, only specific sites (doubly-occupied interstitial or defect sites) are proposed to be "active." Since the number of these active sites is so small, relative to the number of interstitial sites, there must be an additional mechanism to increase their activity levels.

Photo-enhanced Optical Interactions

Next we consider the effect of laser stimulation of LENR. If the laser light is resonantly coupled to the phonon field, the coherence length and amplitude of this field is increased. Thus, we would expect:
1. a higher loading of deuterium into the lattice without going to high temperatures,
2. an increase of the $D^+ D^-$ formation, phasing, and longevity,
3. if properly aligned, an amplification of the breathing mode to allow (or increase) the possibility of direct interaction between deuterons in adjacent interstitial sites.

These actions affect all of the interstitial deuterons, rather than the defect sites (although they can affect doubly-occupied sites). Therefore, light (preferably resonant) has the possibility of enhancing LENR by creating a small effect on a great number of sites.

In the context of light effects on defects, it is appropriate to note the recent discovery of surface-enhanced Raman Scattering[8]. It is believed that the optical interactions are enhanced due to excitations of surface-mode resonances (plasmons, polaritons, etc.). The maximum stimulation takes place at the junction between nanoparticles, at crevices, and at other special locations.

The optical potential at and between the entities (atoms, ions, or molecules) at such locations is given by

$$<U_{op}> = -(1/2)\, \alpha_r\, E_L^2 \qquad (9)$$

where $E_L$ is the electric field acting locally and is the sum of the incident and induced fields resulting from the response of the electrodynamic environment. $\alpha_r$ is the real part of the complex polarizability (taken as isotropic). It will exhibit a peak value of the resonant surface wave at the interface (e.g., PdD-Gold).

Our object of interest is the $D^-$ and $D^+$ pair, the positive ion of which is very mobile in the matrix and on the surface. A strong optical field, by polarization of the charges, increases the probability of a D-D pair being in the $D^- - D^+$ state. The optically-enhanced probability $P_{oe}$ of finding a polarizable species at the point in question can be written as

$$P_{oe} = P_u \{\exp[-U_{op}(l)/k_B T]\}, \qquad (10)$$

where, $P_u$ is the probability of the species being there in the unilluminated (dark) case, and $l$ is the dimension of the region in the interstitial Pd-lattice site aligned with the maximum field (e.g., 1 nm).

In the physical situation of a $D^-$ and $D^+$ pair confined to a palladium interstitial or defect region, the optical field will, hopefully, reduce the free space that the pair can inhabit, from a region of about 3000 nm$^3$ (r = ~10 nm), down to a nearly linear volume of ~ 0.03 nm$^3$ (1nm long, with 0.1 nm radius). This constriction would decrease the interaction volume by a factor of $10^5$. The volume reduction will thereby enhance the nuclear-penetration probability by confining the polar pair to the strongest-field region, by improving the alignment of the collision process, and by increasing the atomic vibration frequency (e.g., from ~$10^{13}$ to >$10^{14}$ sec$^{-1}$). This latter effect would result from the reduction of deuteron recoil length by rebound from lattice atoms determining the gap $l$. Thus, the overall light enhancement could exceed the unilluminated case by <u>more</u> than a factor of $10^5$. [The calculated enhancement could be orders of magnitude higher. However, saturation effects are expected to become important as confinement reduces the average distance of the deuteron pair to the order of the phonon-modified orbitals.]

In the interface region (between PdD-Gold), the near-field evanescent modes will play a dominant role in electric field strengths within the interparticle gaps. The importance of these modes has been seen in our recent studies.[12] For this situation, the optical potential will vary as $C(I, n)(\lambda_o / l)^n$, where $\lambda_o$ is the excitation wavelength in the medium, $C(I, n)$ is a unitless scaling factor incorporating the incident-light intensity, material, and geometrical considerations. The factor n can vary from 2 to 3 (depending upon geometry and spacing). The optical potential U becomes more and more negative with decreasing $l$. In Figure 3, a log-log plot of $P_{oe}/P_u$ vs $l$ exhibits a sharp rise as $l$ diminishes from 5 nm, since $N(l) = - C(I, n)(\lambda_o / l)^n$, a negative number (which can vary from 0 to – 15 over physical dimensions).

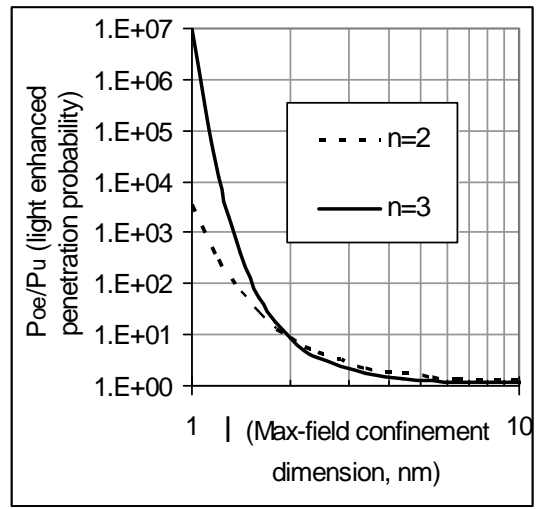

*Figure 3   Ratio of optical-enhancement to unilluminated-penetration probability ($P_{oe}/P_u$) as a function of confinement gap for two different models (see text above).*

To permit comparison of the light-enhancement effect with the bosonic-ion effect, we write the optical potential in terms of temperature factors, $U_{op} = k_B T\, N$. Thus, from equations 5 and 10, the overall bosonic-ion effect plus light-induced enhancement of nuclear interaction per $D^-$ - $D^+$ interaction, $P_E$ becomes

$$P_E = P_T(G') \times (P_{oe}(N)/P_u) = \exp[-G'] \times \exp[|N|] = \exp[-G' + |N|]. \quad (11)$$

When the value of G' is of the order 12 to 17, and |N| is of the order 8 to 15, $P_E$ is enhanced to the point ($P_E = \sim 10^{-4}$) that D-D reactions may be limited by diffusion of new deuterons to the active sites. (Thus, we can see the effect of item 3B in the Theoretical Formulation section.)

As an example of the nuclear reaction <u>rate</u> under these conditions, consider the $D^-$ - $D^+$ confinement to be to r = 1 nm and the atomic vibration frequency to be $10^{14}$ $sec^{-1}$. The effective "flux" (consider the beam flux in a low-energy D-D scattering experiment) would then be $\Phi = \sim (10^{14}/sec) / (3 \times 10^{-14} cm^2) = \sim 3 \times 10^{27} /cm^2$ sec, or a current density within each active site of $\sim 5 \times 10^8$ $A/cm^2$. Multiplying this by the new interaction cross section [from equations 7 and 8, $\sigma = (\pi/k^2)$, where $\pi/k^2 = \sim 2.6 \times 10^{-20}$ $cm^2$, and $P_E = \sim 10^{-4}$] gives $\sim (3 \times 10^{27} /cm^2$ sec$) \times (3 \times 10^{-24} cm^2) = \sim 10^4$ /sec. Thus, fusion of a $D^-$- $D^+$ pair in a single active site would take about 0.1 ms in the modeled light field. Power production thus becomes dependant on the number of active (defect?) sites and the rate at which they can be refilled.

Concluding Remarks:

In the foregoing sections, we have first noted some important experimental results for LENR in $PdD_x$ systems, both in the dark and with laser-stimulated excitation of $PdD_x$-Gold systems. We have presented a model which accounts for both the dark and the illuminated results. The central point is the formation of $D^-$ ions which harbor a pair of electrons in the singlet state at some locations in the solid matrix. These electron pairs screen the nucleus of $D^-$ ions so that a $D^+$ can be attracted and cross the reduced Coulomb barrier. In the laser-stimulated case, the optical potential can enhance both the probability of localized $D^-$ formation and its stability along with the probability of $D^-$ and $D^+$ to come closer and fuse. This effect, along with that of light, greatly increases both bulk and point LENR mechanisms. Therefore, it is still not certain whether the dominant heat production observed under laser illumination would come from deuterons, with low probability of fusion, in very many sites or from deuterons, with a very large probability of fusion, in a much smaller number of sites.

Our model suggests directions to pursue for both cases, so that reproducible and useful results can be obtained in future. For the dark case, increasing the number of defect sites capable of containing deuterium pairs will increase the number of $D^+$- $D^-$ pairs available. Also, determining site structure that will increase the longevity and/or probability of $D^-$ (with a $D^+$ present) would allow material changes to be made that would increase the total power produced from LENR. In the optically-enhanced case, experiments and analysis, to determine the optimal light directions, relative to crystal axes, and the wavelength proximity to resonance frequency for the laser-stimulated experiment, will point to both the validity of the model and then to the selection of materials and laser frequencies needed to improve the results.

Acknowledgements
This work is supported by: HiPi Consulting, New Market, MD, USA; by the Science for Humanity Trust, Bangalore, 560094, India; and the Science for Humanity Trust, Inc, Tucker, GA, USA.